\documentstyle[aps,prl,epsfig]{revtex}
\begin{document}
\draft
\twocolumn[\hsize\textwidth\columnwidth\hsize\csname %
@twocolumnfalse\endcsname

\title{ Mobile Bipolaron}
\author{ $^a$J. Bon\v ca, $^a$T. Katra\v snik, and  $^b$S. A. Trugman}
\address{$^a$ FMF, University of Ljubljana and 
J. Stefan Institute, 1000, Slovenia,\\
$^b$Theory Division,
Los Alamos National Laboratory, Los Alamos, NM  87545,}
\date{\today}
\maketitle
\begin{abstract}\widetext

We explore the properties of the bipolaron in a 1D Holstein-Hubbard
model with dynamical quantum phonons.  Using a recently developed
variational method combined with analytical strong coupling
calculations, we compute correlation functions, effective mass, bipolaron 
isotope effect and the phase diagram.  The two site bipolaron has a
significantly reduced mass and isotope effect
compared to the on-site bipolaron, and is
bound in the strong coupling regime up to 
twice the Hubbard $U$ naively expected.
The model can be described in this regime as an effective
t-J-V model with nearest neighbor repulsion.
These are the most accurate bipolaron calculations to date.

\end{abstract}

]

\narrowtext

While  there  is  a  generally  accepted   belief that  in high  $T_c$
superconductors a dominantly electronic interaction is responsible for
the unusually high  transition temperatures, the interplay between the
electron-phonon   interaction   and   the  strong    electron-electron
interaction nevertheless plays a  significant role in  determining the
physical    properties   of   these    strongly   correlated   systems
\cite{bishop,alex}.  Although the study of lattice effects in strongly
correlated           materials          is        steadily     growing
\cite{grilli,wellein,marsiglio,pao,takada},      the 
understanding of the  influence  of  the Hubbard
interaction  $U$  on  bipolaron  formation  is  still  incomplete.  In
particular, it is known that  in the strong coupling regime bipolarons
have an  extremely large effective mass\cite{kabanov,ranninger}, which
represents one of  the main objections \cite{ranninger,chakra} against
the theory of  small bipolaron superconductivity \cite{alex1}.  Recent
calculations   in  the  adiabatic  (static phonon)
limit    show that    a first-order phase  transition  exists  between the
on-site (S0) and lighter neighboring-site (S1) bipolaron \cite{proville}.  The
properties  of  the   S1   bipolaron  were also  investigated by
variational and exact diagonalization methods \cite{magna,fehske}.

In this letter we present accurate numerical solutions of the 
Holstein-Hubbard model bipolaron on a 1D {\it infinite} lattice. 
Our results are exact to within the line-widths on the figures
in the intermediate and near the strong coupling regimes.
The results are compared to  analytical calculations in the strong
coupling regime.


We consider the Holstein-Hubbard  Hamiltonian \cite{mahan}

\begin{eqnarray}
H && = -t \sum_{js} ( c_{j+1,s}^\dagger c_{j,s} + H.c.)  \label{ham}\\
&& -g\omega \sum_{js} c_{j,s}^\dagger c_{j,s}
( a_j + a_j^\dagger) + \omega \sum_j a_j^\dagger a_j + 
U\sum_{j}n_{j\uparrow}n_{j\downarrow},\nonumber
\end{eqnarray}
where   $c_{j,s}^\dagger$  creates an electron of spin $s$ and
$a_{j}^\dagger$  creates a phonon on  site $j$.     The   last  term in   Eq.~(\ref{ham})  represents the on-site Coulomb repulsion.  
We consider the case
where two electrons with  opposite  spins couple to dispersionless
optical phonons.

Basis states  for the many-body Hilbert  space can be written 
$\vert M \rangle = |j_1,j_2;  \dots ,n_{m}, n_{m+1}, \dots, \rangle$, 
where the up and down electrons are on sites 
$j1$ and $j2$, and there are $n_m$ phonons
on site $m$.  
A variational
subspace is constructed beginning with   an initial state where   both
electrons   are  on the same   site   with no phonons,  and  operating
repeatedly  ($N_h$-times)   with   the off-diagonal   pieces ($t$ and $g$)  
of  the Hamiltonian, Eq.~(\ref{ham}).  All translations of these
states are included on an infinite lattice.
This method was used previously for the polaron (one electron) 
\cite{janez,janez1}.
The method is very
efficient in the intermediate    coupling  regime, where  it   provides
results that are variational in  the thermodynamic limit 
and bipolaron energies
that are accurate to  7 digits for the  case $N_h=18$ and size
of the Hilbert space $N=2.2 \times 10^6$ phonon and down
electron configurations 
for a given up electron position.

Before presenting the numerical calculations, we  show that
many  interesting properties  of the bipolaron can  be found in
second-order    strong  coupling   perturbation  theory.   Following
Lang and Firsov\cite{lang} we use  the canonical transformation $\tilde H
= e^S H e^{-S}$, where $S = g\sum_{js} n_{js}(a_j-a_j^\dagger)$.  The
transformed Hamiltonian takes the form
\begin{eqnarray}
\tilde H &=& H_0 + T  \label{hamtilde}\\
H_0&=&\omega\sum_j  
a_j^\dagger a_j - \omega g^2 \sum_j n_j \nonumber +
\left(U-2\omega g^2\right)\sum_{j}n_{j\uparrow}n_{j\downarrow}\\
T &=& -t e^{-g^2}\sum_{js} \left(c_{j+1,s}^\dagger c_{j,s} 
e^{-g\left(a_{j+1}^\dagger - a_{j}^\dagger\right)} 
e^{ g\left(a_{j+1} - a_{j}\right)} + {\rm H.c.}\right),\nonumber
\end{eqnarray}
where  $n_{j} = n_{j \uparrow} + n_{j \downarrow}$.  
The  first  term in $H_0$ is
the energy  of  the phonon excitations,  and the  second is the energy
gained by displacing the  oscillator in the force of the electron.  
The exponential factors in the $T$ hopping term arise 
because the Lang-Firsov transformation
redefines the origin of the harmonic oscillator
when the number of electrons on the site changes.
In   the limit  $g \to  0$ and  $\omega\to\infty$ with
$\omega g^2$ constant, the phonon interaction
is instantaneous and the
Holstein-Hubbard model maps onto a
Hubbard  model   with an effective  Hubbard interaction
$\tilde  U = U-2\omega g^2$.   In 1D a bound  state  exists for
$\tilde U < 0$.

In the strong coupling or antiadiabatic limit, 
$T$ in Eq.~(\ref{hamtilde}) is considered
a perturbation.  It  represents  the hopping of  electrons,  including
possible  creation and destruction of   phonon excitations.  The  S0
state 
$\phi_0=c^\dagger_{0\uparrow}c^\dagger_{0\downarrow}\vert0\rangle$ has
the lowest energy
to zeroth order   in $T$ when   $\tilde U<0$.
In perturbation theory to second order,
the  energy of the S0
bipolaron is
\begin{eqnarray}
E_{bi}^{S0}(k) &=& U-4\omega g^2 \nonumber \\
&-& {4t^2e^{-2g^2}\over\omega}
\sum_{n= 0}{(2g^2)^n\over n!}{(1+(-1)^n \cos(k))\over 
-\tilde U/\omega + n}.
\label{ebi}
\end{eqnarray}
The effective mass
$m_{S0}^{-1} = \partial^2 E_{bi}(k)/ \partial k^2$ 
can be obtained   from Eq.~(\ref{ebi}) 
by calculating the infinite sums \cite{kabanov}
\begin{equation}
{m_{S0}}^{-1}={4t^2\over\omega}e^{-\left(2g^2-{\tilde U\over\omega} 
\ln 2g^2\right)}\left[\Gamma\left(-{\tilde U\over\omega}\right)-
\Gamma\left(-{\tilde U\over \omega},2g^2\right)\right],
\label{mbi}
\end{equation}
where 	 $\Gamma(x)$  and $\Gamma(a,x)$ are the gamma  and
incomplete gamma functions respectively.
Taking the large $g$ limit in
Eq.~(\ref{mbi}), one finds (see also Ref.~\cite{kabanov})
$m_{SO}^{-1}\simeq 4 \sqrt \pi t^2 \exp(-4g^2) /(\omega g) $ for
$U = 0$.  
At large $g$, $m_{S0}$ is roughly a factor $\exp(3 g^2)$ larger
than the polaron mass, $m_{po}^{-1}\simeq 2 t \exp(-g^2)$. 

One would naively expect that within the strong coupling
approximation, a bipolaron
unbinds when $\tilde U\geq 0$.  This is false: a bound
bipolaron may exist even for $\tilde U\ge 0$.  In this regime, a set
of degenerate states $\phi_i=
c^\dagger_{0\uparrow}c^\dagger_{i\downarrow}\vert0\rangle$ for $i\not
=0$, written in a translationally invariant form, represents states
with minimum energy of $H_0$.  The energy of a S1 bipolaron is
obtained by solving the secular equation $\vert
T_{ij}-\delta_{ij}E_{bi}^{S1}\vert=0$, where matrix elements
$T_{ij}=\langle\phi_i\vert T \vert \phi_j\rangle$ are calculated to
second order in $T$.

The main source of binding arises because
the second order matrix element $T_{1-1}$ for two neighboring electrons
$\mid \uparrow \downarrow \rangle$ to exchange sites
is exponentially larger than
all other off-diagonal matrix elements.  
While in the limit $g\to
\infty$, $T_{1-1}\propto t^2/U$,  the second largest  (first
order) matrix  elements   $T_{i,i+1}\propto t \exp(-g^2)$   for 
$i \neq \{ 0,-1 \}$.        
The exponential suppression arises from phonon rearrangement
whenever the initial and final charge distributions differ.
In         the        singlet        configuration,
$\phi^{S=0}_1=(\phi_1+\phi_{-1})/\sqrt  2$, the diagonal correction to
the energy is given by $T_{11}+T_{1-1}$.  
There is also a contribution
to  $T_{11}$ that  resembles a retardation effect, in which  one electron
hops creating one or more phonon quanta on its initial site,
and  the  second electron   follows absorbing the phonons.   This process,
however, decreases exponentially  with   $g$  as   
$t^2 \exp(-g^2)/(\omega g^2)$,  
and is not strong enough  to bind two polarons in the
triplet   configuration.

To obtain the S1 bipolaron stability criterion, we note that the
secular equation for fixed center of mass momentum $k$ can be mapped
onto a tight-binding model for a linear chain.  Since all the
off-diagonal matrix elements except $T_{1-1}$ scale as $\exp(-g^2)$,
the condition for a bound state is given by $T_{11}+T_{1-1}<T_{ii}$.
Keeping only terms of order $t^2/ ( \omega g^2 )$ and setting $k=0$,
the matrix elements can be expressed as $T_{11}=T_a+T_b$,
$T_{1-1}=T_b$, and $T_{ii}=2T_a;~ i>1$, with
\begin{eqnarray}
T_a&=& {-2t^2e^{-2g^2}\over \omega}
\sum_{n= 1} {(2g^2)^n\over n! n }\sim-{t^2\over \omega g^2},
\label{va}\\
T_b&=& {-2t^2e^{-2g^2}\over \omega}
\sum_{n= 0} {(2g^2)^n\over n! } 
{1\over  {\tilde U\over \omega} + n}\sim-{2t^2\over U },
\label{vb}
\end{eqnarray}
where the final terms refer to the large $g$ limit.
In the large $g$ limit, binding occurs for $U<4\omega g^2$.


The effective S1 bipolaron mass is computed by  approximating 
the S1 bipolaron wavefunction with only
$\phi^{S=0}_1$ (omitting the exponential tail),
\begin{eqnarray}
m_{S1}^{-1}&\simeq&{2t^2e^{-2g^2}\over\omega} \left [ {2\omega\over \tilde U} +
\sum_{n=1} {g^{2n}\over n!}\left ({2\over \tilde U/\omega + n} + 
{1\over n}\right )  \right ] \nonumber \\
&\sim& 2 t^2 e^{-g^2} \left [{2\over U - \omega g^2} + 
{1\over \omega g^2} \right ] \label{ms1}.
\end{eqnarray}
There
are three distinct processes contributing to $m_{S1}$: an S1 pair can
move by one lattice site through an intermediate doubly occupied state
with $n$ phonons (terms that contain $U$ in Eq.~(\ref{ms1})), or
through an intermediate state with only phonon degrees of freedom
(terms without $U$).  

The bipolaron isotope effect is a measure of how the
bipolaron mass varies with the ion mass $M$,
$\alpha \equiv d \ln m_{bi}/ d \ln M $.
This can also be written 
$\alpha = - {1 \over 2} d \ln m_{bi} / d \ln \omega $,
where the derivative is taken with $ \omega g^2 $ held constant.
The bipolaron isotope effect $\alpha$ is equal to
the superconductivity isotope effect
$\alpha_{SC} \equiv -d \ln T_c/ d\ln M$ only
in the low density limit, when superconductivity occurs
as a weakly interacting gas of bipolarons Bose condenses.
Then in mean field theory (ignoring fluctuations that are important in 
low dimensions), the transition temperature $T_c$ at fixed
density is proportional to $m_{bi}^{-1}$, and $\alpha_{SC} = \alpha$.
The bipolaron
isotope effect is expected to change substantially between the S0 and S1
regimes.  Indeed, using Eqs.~(\ref{mbi}, \ref{ms1}) in the
strong coupling limit, we obtain $\alpha_{S0} \sim 2 g^2 - {1 \over 4} $ 
for $U=0$ and
$\alpha_{S0}$ decreasing with increasing $U$.  In the S1 regime,
$\alpha_{S1}\sim g^2/2$ and is only weakly dependent on $U$.
The bipolaron isotope effect has also been calculated by
Alexandrov \cite{alexandrov}; the {\it superconductivity}
isotope effect has been measured by many groups \cite{zhao}.



\begin{figure}[tb]
\begin{center}
\epsfig{file=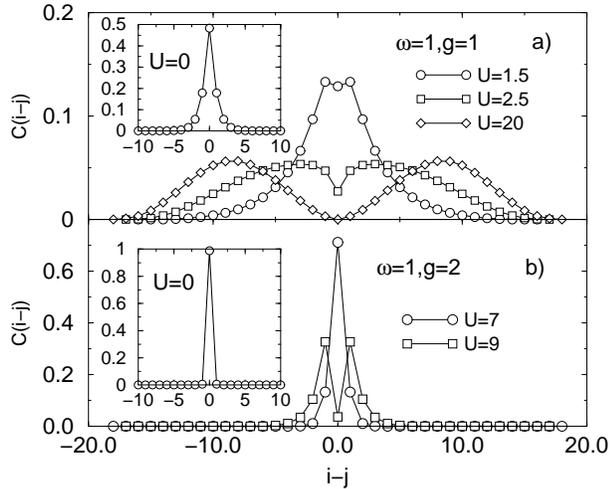,height=80mm,angle=-90}
\end{center}
\caption{Electron-electron correlation function $C(i-j)$ calculated at
$\omega=1$, a) $g=1$    and b)  $g =  2$   for different   values   of
$U$, with $N_h=18$. The two ordinate axes have a different range. Insets show
results for $U=0$. All curves are normalized, $\sum_i  C(i)=1$. }
\label{corel}
\end{figure}

We now present numerical variational results, using units
where  hopping $t=1$.
Fig.~(\ref{corel}) shows the ground state electron-electron density correlation
function $C(i-j)=\langle\psi_o\vert n_i n_j\vert\psi_o\rangle$, where
$n_i=n_{i\uparrow}+n_{i\downarrow}$.
At $g=1$, the bipolaron widens with
increasing $U$ and transforms into two unbound polarons
(which can only move a finite distance apart
in the variational space). The value
$U=1.5$ is below the transition to the unbound state at
$U_c=2.17$, calculated by comparing the polaron and bipolaron energies.
At $U=1.5$, $C(i)$ falls off exponentially, while for $U > U_c$ the typical
distance between electrons is the order of the maximum allowed
separation $N_h$ \cite{max}.
A state of separated polarons is clearly seen for $U=20$.

Two distinct regimes are seen at $g=2$ within the bipolaronic
region.  At $U=7 < U_0 \equiv 2 \omega g^2 = 8$, 
the correlation function represents the S0
bipolaron, while at $U=9>U_0$ we find the largest probability for two
electrons to be on neighboring sites, which is characteristic of
the S1 bipolaron.  In contrast to previous calculations
where phonons were treated classically \cite{proville}, we find 
a crossover rather than a 
phase transition between the two regimes.  
The precision  of  presented correlation  functions  in the bipolaron
regime   is  within the  size  of  the  plot symbols  in the  thermodynamic
limit. 

\begin{figure}[tb]
\begin{center}
\epsfig{file=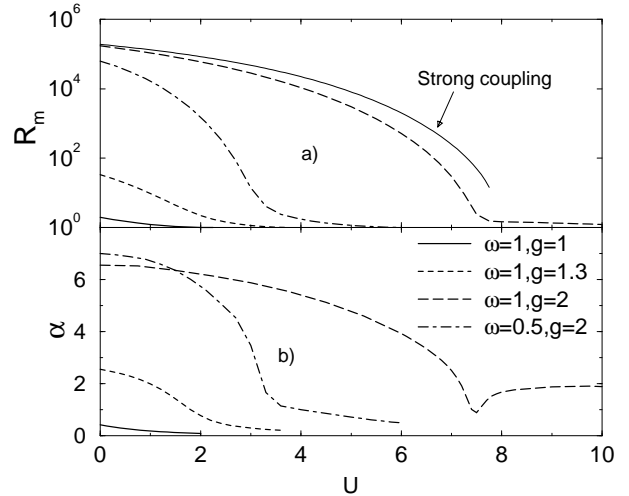,height=80mm,angle=-90}
\end{center}
\caption{a) The mass ratio  
$R_m=m_{bi}/2m_{po}$ vs.  $U$ and b) the bipolaron isotope effect
$\alpha$ vs. $U$.  Numerical results are for $N_h=18$.
Results for $R_m$ at $\omega=0.5$ are obtained 
by extrapolating $N_h \to \infty$.  Precision in all curves is within
the line-width in the
thermodynamic limit, except for
$\alpha$ with $\omega=0.5$, where the error is estimated to be
$\pm 5 \%$. The thin line in a) is 
the strong coupling expansion result for $\omega=1$, $g=2$.  
Polaron masses in units of
the noninteracting electron mass are $m_{po}=1.35, 1.76, 10.4, 3.06$ 
from top to bottom respectively. }
\label{mass}
\end{figure}

Figure (\ref{mass}a) plots the  bipolaron  mass
ratio $R_m=m_{bi}/2m_{po}$  vs.  $U$   for different  values   of
$\omega$  and  $g$.  
In all cases presented in Fig.~(\ref{mass}),
$R_m$ approaches 1 as $U$ approaches $U=U_c$ in agreement with a
state of  two free polarons.  At  fixed  $\omega=1$ the bipolaron mass ratio
increases by several orders of magnitude with increasing $g$ at $U=0$.
The increase can be understood within strong  coupling  theory. 
Increasing $U$ sharply decreases $R_m$ in the S0 regime.
Note that the scale in Fig.~(\ref{mass}) is
logarithmic.  Near the strong coupling regime $(g =2)$ and for small
$U$, good agreement is found between the numerical and the strong
coupling result obtained from Eq.~(\ref{mbi}).  The difference
between these results increases as $U$ approaches $U_0=8$, where the
perturbation theory based on the S0 bipolaron breaks down.  In the S1
regime $U>U_0$, $R_m$ is small, as predicted by the strong
coupling result.

The bipolaron isotope effect, shown in Fig.~(\ref{mass}b), is large in the
strong coupling ($\omega=1,g=2$) and small $U$ regime, where its value
is somewhat below the large $g$ strong coupling prediction 
$\alpha_{S0}\sim 2 g^2 - {1 \over 4} = 7.75$.  
With increasing $U$, $\alpha$ decreases and in the S1 regime
approaches $\alpha_{S1}=g^2/2= 2$. A kink is observed in the crossover
regime. 

\begin{figure}[tb]
\begin{center}
\epsfig{file=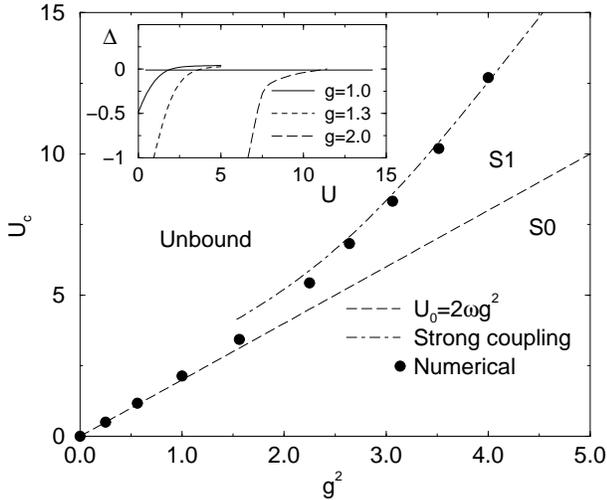,height=80mm,angle=-90}
\end{center}
\caption{Phase diagram and binding energy $\Delta$ in units of $t$ 
(inset) calculated  at
$\omega=1$.  Numerical results
are circles.  For  greater accuracy, results near
the weak and strong  coupling regime were obtained 
by extrapolating $N_h \rightarrow \infty$.}
\label{diag}
\end{figure}

We conclude with the  phase     diagram   $U_c(g)$  shown in
Fig.~(\ref{diag}) at fixed $\omega = 1$.  
Numerical results, shown as circles, indicate the phase boundary
between two dissociated polarons each having energy $E_{po}$ and a
bipolaron bound state with energy $E_{bi}$. In the inset of
Fig.~(\ref{diag}) we show the bipolaron binding energy  defined
as $\Delta = E_{bi}-2 E_{po}$. The phase diagram is obtained from $\Delta=0$.
The dashed line,
given by  $U_0=2\omega g^2$, is a  reasonable estimate for the
phase  boundary at small $g$.   At large  $g$  the dashed line roughly
represents  the crossover between a
massive S0 and lighter S1 bipolaron.  
The   dot-dashed  line is the phase   boundary
between  S1 and the unbound polaron phase,   as  obtained by
degenerate  strong  coupling  perturbation theory.  Numerical  results
approach  this line at  larger  $g$.  The  dot-dashed  line asymptotically
approaches $U_c=4\omega g^2$.

In the S1 regime, the strong coupling expansion
can be used to rewrite the 
Holstein-Hubbard Hamiltonian  Eqs.~(\ref{ham},\ref{hamtilde})
as an effective t-J-V model that applies to
an arbitrary number of particles, 
\begin{eqnarray}
H_{eff} &=& -t_1\sum_{i,s}\tilde c^\dagger_{i,s}\tilde c_{i+1,s}
+h.c. + \sum_i J{{\bf S}_i}{{\bf S}_{i+1}} + V n_in_{i+1}, \nonumber
\end{eqnarray}
where $t_1=t\exp(-g^2)$, $J =  -2T_b>0$, $V=-T_a+T_b/2>0$ and  $\tilde
c_{i,s}=   c_{i,s}(1-n_{i,-s})$ \cite{hotta}.  The $V$ term is a repulsion
between nearest neighbors.
For simplicity  and in keeping with the approximations used 
to derive the standard $t-J$ model, we  have  omitted
next-nearest neighbor hopping terms that  are of order 
$t^2 \exp(-g^2) /(\omega g^2)$, and a constant  term.    
While the standard $t-J$ model can be derived from the
Hubbard model only for $J < t$, the
parameters of  $H_{eff}$ are quite different,
with $J$ typically much larger than $t_1$.
In  the static limit, an
S1 bipolaron is stable in the  interval $J/4<V<3J/4$.
In the dilute
limit, only  singlet bipolarons exist with  binding energy (in the
static limit) $E_{S=0}=-3J/4+V\sim-4t^2/U+t^2/ (\omega  g^2)$,  while
triplet fluctuations are almost  completely frozen  out
due  to the  high energy $E_{S=1}=J/4+V$.
Such a   state is very  similar to  a
negative-$U$ Hubbard model, except that singlets occupy
two sites, like in the RVB model.

In conclusion, we have demonstrated that near the strong coupling limit
a mobile S1 bipolaron exists with an effective mass of the order of a
polaron mass, and an isotope effect in the range
$0<\alpha<g^2/2$.  The wavefunction of the S1 bipolaron is a spin
singlet with extended $s-$wave spatial symmetry.  Taking into
account the asymptotic stability criterion $U_c=4\omega g^2$, it is
clear that a triplet S1 bipolaron that corresponds to the $U\to\infty$
solution is not bound.  In the static limit, it can be shown
that bound states of three or more polarons are not stable in the S1
regime, thus ruling out  phase separation.  An effective t-J-V
model captures the physics of many S1 bipolarons in the strong coupling
regime of the Holstein-Hubbard model, where antiferromagnetism is a consequence
of the original Hubbard interaction $U$, while the attractive interaction is
mediated by phonons. Taking into account the similarity between a system
of S1 bipolarons and the negative-U Hubbard model, one should not rule
out the possibility that S1 bipolarons form a superconducting state of
either $s-$wave or $d-$wave symmetry in two spatial dimensions.  In
such a state,  strong electron-electron interactions and
electron-phonon coupling should be  treated on an equal footing.

We acknowledge stimulating discussions with V.V. Kabanov and F. Marsiglio.
J.B.  greatfully  acknowledges the    support of Los   Alamos National
Laboratory   where part  of  this   work has  been   performed, and the
financial support by the Slovene Ministry of Science.
This work was supported in part by the US DOE.


\begin{references}
\bibitem{max}  The electrons can be
no farther apart than $N_h$ in the variational space, 
although their center of mass
can be anywhere on an infinite lattice.
\end{references}
\end{document}